\documentclass[twocolumn,prl,showpacs]{revtex4}
\usepackage{epsfig}
\usepackage{graphicx}
\usepackage{psfig,epsfig}
\usepackage{dcolumn}
\usepackage{bm}

\def\e{\mbox{e}}

\begin{document}

\title{Segregation in fluidized versus tapped packs}
\author{Marco Tarzia $^{a}$, Annalisa Fierro $^{a,b}$, Mario Nicodemi $^{a,b}$, 
Antonio Coniglio $^{a,b}$}
\affiliation{${}^a$ Dipartimento di Scienze Fisiche, Universit\`{a} degli Studi
di Napoli ``Federico II'', INFM and INFN, via Cinthia, 80126 Napoli, Italy}
 
\affiliation{${}^b$ INFM - Coherentia, Napoli, Italy}                           
\date{\today}

\begin{abstract}
We compare the predictions of two different statistical mechanics approaches,
corresponding to different physical measurements,
proposed to describe binary granular mixtures 
subjected to some external driving
(continuous shaking or tap dynamics). In particular we analytically solve 
at a mean field level the partition function of
a simple hard sphere lattice model under gravity and we
focus on the phenomenon of size segregation.
We find that the two approaches lead to similar results and seem to coincide
in the limit of very low shaking amplitude. However they give
different predictions of 
the crossovers from Brazil nut effect to reverse Brazil nut effect with 
respect to the shaking amplitude, which
could be detected experimentally.
\end{abstract}

\maketitle
Segregation of hard sphere mixtures is a relevant and historically debated
problem (see \cite{Evans} and ref.s therein). The phase behavior
of these systems is still hotly debated and it is still controversial
whether or not hard sphere mixtures segregate in absence of 
gravity. 
In the last decade a great attention has been devoted to the study 
of the problem of vertically shaken granular mixtures under gravity.
It was observed that such systems
can mix or, under different
conditions,  segregate their components
spontaneously according to criteria which are still largely unclear, 
although of deep
practical and conceptual relevance \cite{rev_segr,capri,nagel}.
Rosato et al. \cite{rosato} demonstrated via molecular dynamics simulations 
that in some cases large spheres
segregate to the top and small spheres to the bottom of the container
when subjected to shaking.
This phenomenon is commonly called Brazil Nut Effect (BNE) 
(while the opposite
one, i.e. large spheres on the bottom and small ones on the top, is known
as Reverse Brazil Nut Effect (RBNE) \cite{luding,breu}).
The authors suggested an interpretation of the BNE as a 
geometric effect (called ``percolation'') where small grains pass through
the holes created by the larger ones \cite{rosato,bridgewater}.
Along with geometry, dynamical effects associated with grains,
such as inertia \cite{shinbrot} or convection \cite{knight}, which bring
large particle up but does not allow them to re-enter in downstream,
were also shown to play a role \cite{jenkins}.

Recent results have
however outlined that segregation processes can involve ``global''
mechanisms such as ``condensation'' \cite{luding} or
phase separation \cite{mullin, seg}.
This suggested a change of perspective on the issue and
the idea to formulate a statistical mechanics description of 
these phenomena.
In Ref.s \cite{luding, both} a statistical mechanics approach was proposed to 
describe segregation in granular mixtures.
Assuming that, as in standard thermal systems, 
the grain kinetic energy may play the role of the bath temperature, $T_{bath}$,
the system was approximated as a standard fluidized gas of elastic hard spheres
kept at a given temperature, $T_{bath}$, dependent on the shaking amplitude. 
This approach may be appropriate to describe segregation phenomena in 
granular mixtures under continuous shaking, but  not systems  subjected to 
a ``tap dynamics'' (where the energy is pumped into
the system in pulses, and the measures are performed when the system is at
rest).  
Following the idea, originally
suggested by Edwards \cite{Edwards1}, and recently further studied \cite{e1}
(see \cite{capri} for a review), it is possible to
develop a statistical mechanics description of granular mixtures under such a 
dynamics.
Edwards' assumption consists in the hypothesis that in a granular
system under taps, time averages 
coincide with suitable ensemble averages  over the ``mechanically stable''
states, i.e., those where the system is found still.
Recently, it was shown \cite{nfc} that this statistical mechanics 
approach holds in good approximation for 
a schematic lattice model of a hard sphere binary mixture under gravity.
In this case two ``configurational'' temperatures  have to be 
introduced in order to describe the system's macroscopic states: These 
temperatures  are just the inverse thermodynamic parameters canonically 
conjugate to the  gravitational energies of the two species,  
and are not related to the grain kinetic energy which is always zero. 

In the present paper we study the relation between these two different 
approaches in the framework of a simple lattice model for 
hard sphere binary mixtures under gravity. 
In particular we analytically solve, at the level of 
Bethe approximation, the partition function of the system
in two different cases 
(which we will refer to as case I and case II): I) All the 
configurations are allowed, as in a gas (the system is
treated as a standard fluidized hard sphere gas under gravity, and it is 
not at rest); II) in the framework of Edwards' theory, where the particles are 
required to be in a stable configuration (system at rest).
We derive the mixing/segregation properties as a function of various
parameters such as grain masses, sizes, numbers and others, 
and finally we compare the results obtained in the two cases.
We find that in both cases the system moves from BNE to RBNE
increasing the mass ratio $m_2/m_1$ or decreasing 
the concentration ratio $N_2/N_1$ of the two species, instead
opposite behaviors are found  increasing the shaking amplitudes 
(i.e. the temperatures). 
In particular in case I the system
moves from a BNE state to a RBNE increasing the shaking amplitude,
and in case II the opposite one is found. 
Moreover, for some values of the mass ratio $m_2/m_1$,  
the RBNE is always found (except at very low temperatures) in the case I,
while the BNE is always found in the case II. 
This scenario could be experimentally checked.

The model we consider \cite{nfc,seg} is a hard sphere binary mixture
made up of two species, 1 (small) and 2 (large) with grain diameters $a_0=1$
and $\sqrt{2} a_0$, under gravity on a cubic lattice confined in a rigid box.
On each site of the lattice we define an occupancy variable, $n_i^z=0,1,2$, 
respectively if
site $i$ at height $z$ is empty, filled by a small or by a large grain.
The Hamiltonian is:
${\cal H}={\cal H}_{HC}+ m_1gH_1 + m_2gH_2$,
where $H_1=\sum_{i,z} z \, \delta_{n_i^z 1}$,
$H_2=\sum_{i,z} z \,\delta_{n_i^z 2} $ are the heights of the two species,
and ${\cal H}_{HC}$ is the hard core potential, preventing two nearest
neighbor sites to be both occupied if at least one contains a large grain.

The partition function of the system
for the case of the fluidized hard sphere gas subjected to continuous shaking
(case I) is given by:
\begin{equation}
\nonumber
{\cal Z}_I =  \sum_{\{r\}} \e^{-\left[{\cal H}_{HC} (r) + \beta_{bath}
\Big( m_1gH_1 + m_2gH_2 \Big) \right]},
\label{Z1}
\end{equation}
where the sum is over all microstates $r$ and $\beta_{bath}$ 
is the inverse bath temperature. 

As shown in \cite{nfc} the system subjected to tap dynamics
is well described by Edwards' approach.
In this case the weight of a given state $r$ is \cite{nfc}:
$\exp\left\{-{\cal H}_{HC}(r)-\beta_1m_1gH_1(r) 
-\beta_2m_2gH_2(r)\right\}{\cdot} \Pi_r$,
where $\beta_1$ and $\beta_2$ are the variables conjugate respectively to the
gravitational energies of the two species.                                      
The operator $\Pi_r$ selects mechanically stable states:
$\Pi_r=1$ if $r$ is ``stable'', else $\Pi_r=0$.
We adopt a simple definition of ``mechanical
stability'': a grain is ``stable'' if it has a grain underneath.
For a given grain configuration, $r=\{n_i\}$, the operator $\Pi_r$ has a
tractable expression:
$\Pi_r =\lim_{K\rightarrow\infty}\exp\left\{-K{\cal H}_{Edw}\right\}$
where ${\cal H}_{Edw}=
\sum_{i,z} \left[ \delta_{n_i^z 2}\delta_{n_i^{z-1} 0}\delta_{n_i^{z-2} 0}
+ \delta_{n_i^z 1}\delta_{n_i^{z-1} 0}
\left(1 - \delta_{n_i^{z-2} 2}\right) \right]$.                                 

The system partition function in this case is given by:
\begin{equation}
\nonumber
{\cal Z}_{II} =  \sum_{\{r\}} \e^{-\left[{\cal H}_{HC} (r) + \beta_1
m_1gH_1 + \beta_2 m_2gH_2 \right]}
{\cdot} \Pi_r,
\label{Zed}
\end{equation}
where the sum is again over all microstates $r$ but, due to the projector,
only the mechanically stable ones are taken into account.
We found that the configurational temperatures,
$T^{conf}_1 \equiv \beta_1^{-1}$ and $T^{conf}_2\equiv \beta_2^{-1}$,  
increase as function of the tap amplitude, and tend to coincide in the
limit of low tap amplitudes \cite{prep}. Thus for simplicity in the following 
we put $\beta_{conf} \equiv \beta_1= \beta_2$.

\begin{figure}[ht]
\vspace{-0.15cm}
\begin{center}
\psfig{figure=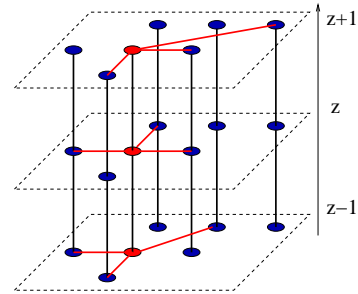,width=4.6cm,angle=0}
\end{center}
\vspace{-0.55cm}
\caption{Bethe lattice used in the calculation}
\label{randomgraph}
\vspace{-0.25cm}
\end{figure}

Since the exact calculation of ${\cal Z}_I$ and ${\cal Z}_{II}$
is hardly feasible, we evaluate the partition functions
at a mean field level. To this aim
we consider a generalization of Bethe-Peierls method for anisotropic systems 
(due to gravity) already used in previous papers \cite{cdfnt,seg}, 
i.e., we solve the
partition function of the system on the Bethe lattice shown in  
Fig.\ref{randomgraph} by means of recurrence
relations (see also \cite{MP}). In particular, we consider
a $3D$ lattice box with $H$ horizontal layers (i.e., $z\in\{1,...,H\}$)
occupied by hard spheres. Each layer is a random graph of given
connectivity, $k-1$ (we take $k=5$). Each site in layer $z$ is also
connected to its homologous site in $z-1$ and $z+1$
(the total connectivity is thus $k+1$).
The Hamiltonians are the ones above 
plus two chemical potential terms which control the two species concentrations.
Hard core repulsion prevents two particles on connected sites to overlap.

In the following we give the main ideas of the calculation in the case I, and 
refer to a longer paper \cite{prep} in preparation for the details, and the 
calculation in the case II (see also \cite{seg, cdfnt}). 
The lattice of Fig. \ref{randomgraph} is iterated in three different 
directions, 
``side'', ``up'' and ``down''. As consequence three different branches exist,
and the corresponding partition functions, $Z_{0,(s,u,d)}^{(i,z)}$ and
$Z_{n,(s,u,d)}^{(i,z)}$ (restricted respectively to configurations
in which the site $i$ is empty or filled by a particle of species $n=1,2$) 
can be defined. The Bethe-Peierls recursion equations, which allow calculation
of the partition function, are more easily written in terms of  
the local ``cavity fields'' defined by:
$e^{h_{n}^{(i,z)}} = Z_{n,s}^{(i,z)}/Z_{0,s}^{(i,z)}$,
$e^{g_{n}^{(i,z)}} = Z_{n,u}^{(i,z)}/Z_{0,u}^{(i,z)}$,
$e^{f_{n}^{(i,z)}} = Z_{n,d}^{(i,z)}/Z_{0,d}^{(i,z)}$,
(with $n=1,2$).               

The fluid phase corresponds to a solution of Bethe-Peierls equations
where local fields in each layer are site independent \cite{nota_crys}.
Such a solution, characterized by horizontal translational invariance, is
given by the fixed points of the following equations:
\begin{eqnarray} \label{ricorrenza}
\nonumber
e^{h_{n}^{(z)}} & = & e^{\beta_{bath}(\mu_n-m_ngz)}
\left[
\frac{A_n^{(z)}}{S^{(z)}} \right]^{k-2}
\frac{U_n^{(z+1)}}{P^{(z+1)}}\frac{D_n^{(z-1)}}{Q^{(z-1)}},
\\
e^{g_{n}^{(z)}} & = & e^{\beta_{bath}(\mu_n-m_ngz)}
\left[
\frac{A_n^{(z)}}{S^{(z)}}\right]^{k-1} \frac{U_n^{(z+1)}}{P^{(z+1)}},
\\
\nonumber
e^{f_{n}^{(z)}} & = & e^{\beta_{bath}(\mu_n-m_n gz)}
\left[
\frac{A_n^{(z)}}{S^{(z)}}\right]^{k-1} \frac{D_n^{(z-1)}}{Q^{(z-1)}},
\end{eqnarray}
where $n=1,2$; $\mu_{1,2}$ are the chemical potentials of the two species
and $m_{1,2}$ are the grain masses;
$A_n^{(z)} = 1+\delta_{n1} e^{h_{1}^{(z)}}$,
$S^{(z)}=1+e^{h_{1}^{(z)}}+e^{h_{2}^{(z)}}$,
$U_n^{(z)} = 1+ \delta_{n1} e^{g_{1}^{(z)}}$,
$D_n^{(z)} = 1+ \delta_{n1} e^{f_{1}^{(z)}}$, 
$P^{(z)} = 1+e^{g_{1}^{(z)}}+e^{g_{2}^{(z)}}$ and
$Q^{(z)} = 1+e^{f_{1}^{(z)}}+e^{f_{2}^{(z)}}$.
From the local field the system free energy can be derived: 
\begin{equation} \label{free_energy}
F = \sum_{z=0}^H \Delta F_s^{(z)} - \frac{(k-1)}{2} \sum_{z=0}^H \Delta F_{l,2}^{(z)} 
- \sum_{z=0}^{H-1} \Delta F_{l,1}^{(z)},
\end{equation}
where,
\begin{eqnarray} \label{eq:shift}
e^{- \Delta F_{s}^{(z)}} & = & \left(S^{(z)}\right)^{k-1}
P^{(z+1)} Q^{(z-1)} \\
\nonumber
&+& \sum_{n=1,2} e^{\beta_{bath}(\mu_n-m_ngz)} 
\left( A_n^{(z)} \right)^{k-1} U_n^{(z+1)} D_n^{(z-1)}, \\
\nonumber
e^{-\Delta F_{l,1}^{(z)}} & = & 
P^{(z+1)} +  \sum_{n=1,2} e^{f_n^{(z)}} U_n^{(z+1)},\\
\nonumber
e^{-\Delta F_{l,2}^{(z)}} & = & 
1 + 2 e^{h_1^{(z)}} +  2 e^{h_2^{(z)}} + e^{2 h_1^{(z)}}.
\end{eqnarray}
A similar calculation can be developed in case II.
In the following instead of using the chemical potential variables, $\mu_1$ 
and $\mu_2$, we will use the conjugate variables, $N_1$ and $N_2$,
the number per unit surface respectively of the small and large grains. 
                       
\begin{figure}[ht]
\vspace{-0.5cm}
\begin{center}
\psfig{figure=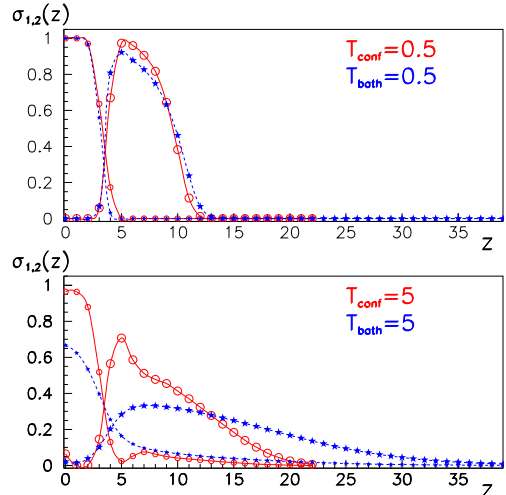,width=7.3cm}
\end{center}
\vspace{-1.1cm}
\caption{(color online). Density profiles in case I, where all the 
configurations are allowed 
(blue stars and dashed lines), and in case II, where  only the stable 
configurations are allowed (red empty circles and continuous lines) for 
$m_1=m_2=1$ $N_1=3.6$, $N_2=2.8$, and respectively $T_{conf}=0.5$ and 
$T_{bath}=0.5$ (top), and $T_{conf}=5$ and $T_{bath}=5$ (bottom). In all the
cases small grains are on the box bottom (BNE).}
\label{profiles}
\vspace{-0.15cm}
\end{figure}

From the free energy $F$, we calculate 
the density profiles, $\sigma_{1,2}(z)$, defined by: 
\begin{equation}
\sigma_n(z)=\frac{1}{N}\left(\frac{2 r_n}{a_0}\right)^2 \sum_i n_i(z)
\delta_{n n_i}
\end{equation} 
where $N$ is the site number on each layer, $n=1, 2$ corresponds respectively 
to small and large grains, and $r_1$ and $r_2$ are respectively the radius of 
small grains ($a_0/2$) and of large ones ($\sqrt{2}a_0/2$).                     In Fig. \ref{profiles} we compare $\sigma_{1,2}(z)$
obtained in the two cases for low and high temperatures.                                 
At low temperatures the density profiles
are quite similar, as expected since, in the case I, the system
explores almost exclusively the mechanically stable states.
At high temperatures the density profiles are instead much different.

\begin{figure}[ht]
\vspace{-0.65cm}
\begin{center}
\psfig{figure=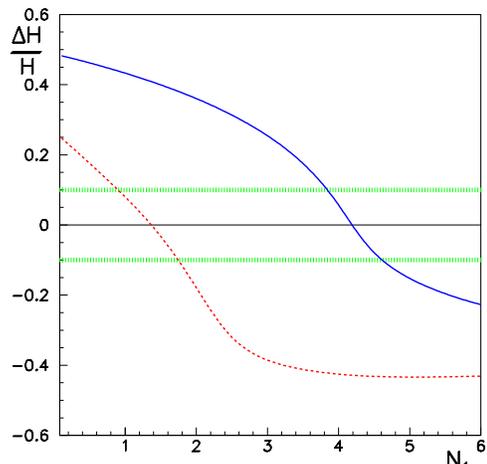,scale=0.35}
\end{center}
\vspace{-1.0cm}
\caption{(color online). $\Delta H/H$ as a function of $N_1$ (for $N_2=0.4$) in 
the case I (blue continuous line on the top) for $T_{bath}=2$,
and in the case II (red dashed line on the bottom) for $T_{conf}=2$. The broad
horizontal lines identify the ``mixing region'' where $-0.1 < \Delta H / H <
0.1$.} 
\label{enne}
\vspace{-0.15cm}
\end{figure}

In Fig. \ref{enne}, $\Delta H / H = (H_1 - H_2)/(H_1 + H_2)$ is plotted as a 
function of the concentration, $N_1$, of the small grains, at
a given concentration of the large ones, $N_2$: In both cases
the system shows a crossover from BNE to RBNE as 
the concentration is decreased.

\begin{figure}[ht]
\vspace{-0.45cm}
\begin{center}
\psfig{figure=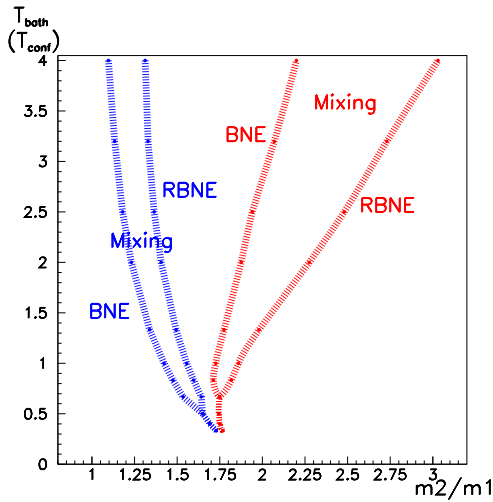,scale=0.34}
\end{center}
\vspace{-0.55cm}
\caption{(color online). Diagram of the vertical segregation state in the plane 
mass ratio - temperature (respectively $T_{bath}$ and $T_{conf}$) for
$N_1 = 0.8$ and $N_2=0.4$. The blue lines on the left are relative to the 
case I, and the red lines on the right are relative to the case II.}
\label{confronto}
\vspace{-0.1cm}
\end{figure}

In Fig. \ref{confronto} a diagram of the vertical segregation is given
in the plane mass ratio - temperature (respectively $T_{bath}$ and $T_{conf}$ in
the two cases) for fixed values of $N_1$ and $N_2$.
The broad lines correspond to the crossover from BNE to ``mixing'', and
from ``mixing'' to RBNE  (the ``mixing region'' is defined as $-0.1 < \Delta H 
/ H < 0.1$). The lines on the left are those 
relative to the case I, which approach asymptotically
the vertical line $m_2/m_1=1$ in the limit $T_{bath}\rightarrow\infty$, as in
a point-like bidisperse gas under gravity. 
At high temperatures in both cases the system moves from BNE to RBNE with a 
continuous crossover. Increasing the temperatures the mixing regions broaden 
since the entropic term in the free energy becomes more important, and the
curves in the two cases get apart since the differences of the entropic terms 
increase. At very low temperatures 
in both cases the mixing region becomes a line  due to the presence
of a real phase transition in the corresponding
model without gravity \cite{seg}: small and
large grains tend to demix and due to the gravity one of the two species is
energetically favored to stay on the bottom depending on the mass ratio.
By further decreasing the temperatures the curves in the two cases become 
closer, and finally tend to a common value in the limit $T \to 0$.  

It is evident that at given mass ratio the
crossover from BNE to RBNE is obtained by varying the shaking amplitude 
(i.e. the temperatures) in opposite ways
in the two cases: In the case I a crossover from BNE to RBNE is found
when $T_{bath}$ is increased as experimentally 
observed in \cite{breu}, and in the case II a crossover from RBNE to BNE is 
found when $T_{conf}$  is increased.
Moreover, for some values of the mass ratio,
the RBNE is always found (except at very low temperatures) in the case I,
instead the BNE is always found in the case II.
This scenario could be checked via molecular dynamics simulations and 
experiments of binary granular mixtures:                                  
Increasing the shaking amplitude of
the continuous external shaking (case I) and of the tap dynamics (case II)
in fact corresponds respectively to increasing $T_{bath}$ and $T_{conf}$.       

In conclusion we focus on the problem of the vertical size
segregation in binary granular mixtures subjected to external driving.
We compare the predictions of two statistical
mechanics approaches proposed to describe granular materials, the 
first more appropriate to deal with fluidized systems subjected to
continuous shaking and the second with granular materials subjected
to tap dynamics. Applying the two approaches to a simple hard sphere
lattice model under gravity we find, in contrast to similar features at low
shaking amplitudes, very different behaviors (which could be experimentally
checked) as the shaking amplitude is increased. This is due to the prevailing 
of two different mechanisms giving rise to segregation: for the fluidized
granular mixture the segregation is mainly energetic driven, instead for the
tapped mixture the segregation is mainly entropic driven. A consequence is for
example that the BN region in the second case is larger than in the first one,
due to the fact that the stable configurations with small grains 
on the box bottom are more than those with large grains on the bottom. 

Work supported by EU Network Number  MRTN-CT-2003-504712, MIUR-PRIN 2002, 
MIUR-FIRB 2002, CRdC-AMRA, INFM-PCI.


\begin{thebibliography}{40}
\bibitem{nota_crys}
Along with the fluid solution, also a crystalline solution,
characterized by the breakdown of the horizontal translational
invariance, can be found. Here we consider the case where the crystallization
is avoided, and limit the calculation to the fluid phase.
\end{thebibliography}
\end{document}